\documentclass[aps,prd,amssymb,cite,
amsfonts,epsf,preprintnumbers,nofootinbib,superscriptaddress]{revtex4}

\usepackage{amsmath,amssymb,amsthm}
\usepackage{hyperref}
\hypersetup{colorlinks=true,linkcolor=blue,citecolor=blue,urlcolor=blue}
\usepackage{bm}
\usepackage[dvips]{graphicx}
\usepackage{subcaption}
\usepackage{latexsym,amsfonts}
\usepackage[usenames,dvipsnames]{color}
\usepackage{soul}
\soulregister\cite7
\soulregister\ref7
\soulregister\pageref7
\usepackage{epsfig}
\usepackage{braket}
\usepackage[mathscr]{eucal}
\usepackage{cancel}
\usepackage{mathrsfs}
\usepackage{pgf, tikz}
\usepackage{slashed}
\usetikzlibrary{arrows,automata}
\usepackage{pgfplots}
\pgfplotsset{compat=newest}

\usepackage{amsthm}

\newtheorem{proposition}{Proposition}

\theoremstyle{remark}

\newcommand{\be}[1]{\begin{equation} \label{#1}}
\newcommand{\ee}{\end{equation}}
\newcommand{\bea}{\begin{eqnarray}}
\newcommand{\bean}{\begin{eqnarray*}}
\newcommand{\eea}{\end{eqnarray}}
\newcommand{\eean}{\end{eqnarray*}}
\newcommand{\ba}{\begin{array}}
\newcommand{\ea}{\end{array}}
\newcommand{\bel}{\begin{align}}
\newcommand{\eel}{\end{align}}

\newcommand{\qarrow}{\quad\rightarrow\quad}

\renewcommand{\d}{\mathrm{d}}

\begin{document}

\title{Collective modes and screening in an electric-magnetic dual plasma}
\author{Hyeong-Chan Kim}
\affiliation{School of Liberal Arts and Sciences, Korea National University of Transportation, Chungju 380-702, Korea}
\email{hckim@ut.ac.kr}
\date{\today}

\begin{abstract}

We study the linear response of an effective relativistic two-fluid medium
carrying separately conserved electric and magnetic charge currents.  
The model is defined by the duality-symmetric Maxwell equations with electric and magnetic sources, together with Lorentz-force dynamics for two fluids with independent inertia and possible Carter-type entrainment.  
The magnetic component is treated as an effective charge-carrying constituent, so the analysis uses only the closed two-fluid equations.

Around a homogeneous, neutral, and unmagnetized background, the transverse electromagnetic response contains two stable branches whose cutoffs are set by the electric and magnetic plasma frequencies and are exchanged by electric--magnetic duality.  
In the longitudinal sector, entrainment mixes the electric and magnetic density oscillations, turns their crossing into an avoided crossing, and gives the stability condition $  \kappa^2<1 ,$ equivalent to positive definiteness of the two-fluid momentum matrix.
Resolving the magnetic component into monopole and antimonopole species gives a neutral branch selected by magnetic charge conjugation \(C_m\).  
In this branch the net magnetic current vanishes, so the long-range monopole field is absent, while the total magnetic density can still produce screened collective response.  
The resulting picture is that magnetic charge can be statically hidden but dynamically visible.  A robust observable signature is the density scaling
$
        \omega_{\rm coll}^2\sim\omega_{pm}^2\propto n^0_{(m)} ,
$
which may survive dissipative broadening even when sharp ideal-plasma poles are not resolved.
We briefly comment on possible dyonic interpretations of magnetically
neutral composites, but the linear-response results do not rely on that
interpretation.

\end{abstract}

\maketitle

\section{Introduction}
\label{sec:intro}

Magnetic charge is usually discussed as an additional source of the Maxwell
field.  In the standard Dirac picture, a magnetic monopole is represented by a
singular potential, and the unobservability of the Dirac string leads to charge
quantization~\cite{Dirac1931,Dirac1948,Schwinger1966}.  The same physics can be described
globally using gauge patches in the Wu--Yang formulation~\cite{WuYang}.
Equivalent local descriptions, such as the two-potential formulation of
Cabibbo and Ferrari, the local Zwanziger action, and the covariant PST
construction, make electric--magnetic duality manifest at the price of an
enlarged potential structure~\cite{CabibboFerrari,Zwanziger,PST}.  These
formulations establish how magnetic charge can be included consistently in
electrodynamics.  The question addressed here is different and more
phenomenological: if magnetic charge is carried by a dynamical medium with its
own inertia, what collective electromagnetic response follows?

We therefore study an effective electric--magnetic two-fluid plasma.  The
minimal closed model consists of the duality-symmetric Maxwell equations with
electric and magnetic sources,
\[
        \nabla_b F^{ab}=j^a_{(e)},\qquad
        \nabla_b{}^*F^{ab}=j^a_{(m)}.
\]
The two charge currents are carried by independently conserved electric and magnetic fluids, with charge densities \(j^a_{(e)}=q n^a_{(e)}\) and \(j^a_{(m)}=g n^a_{(m)}\). Their inertial response, including possible Carter-type entrainment, is specified in Sec.~\ref{sec:setup}. No microscopic monopole core model is assumed; the magnetic component is treated as an effective charge-carrying constituent whose inertia is a low-energy parameter of the medium.

One motivation for this effective system comes from the convective
variational, or matter-space, description of relativistic fluids and
multifluids developed by Carter and subsequent authors
~\cite{Carter,CarterMultifluid,CarterLanglois,AnderssonComer2021,Kim:2022qlc},
and from its application to electromagnetism and two-sector
decompositions~\cite{Ho2025,matterspaceEM}.  In that framework conserved
currents arise naturally from matter-space pull-backs, and a magnetic
charge-carrying constituent can be represented at the effective level by a
conserved magnetic current.  The present paper does not require the details of
that construction: it takes the closed electric--magnetic two-fluid equations
as an effective model and computes their linear collective response.

We linearize the dual plasma around a static, homogeneous, neutral, and
unmagnetized background.  The central question is whether the magnetic carrier
is dynamically inert when the background has no net monopole field, or whether
it leaves observable signatures in the collective spectrum.  The answer is
that a magnetically neutral medium can be statically hidden but dynamically
visible.  In the transverse sector the response contains two duality-related
plasma scales, set separately by the electric and magnetic plasma frequencies.
The resulting dispersion relation has two stable branches; when the magnetic
plasma frequency is taken to zero, one branch reduces to the ordinary electric
plasma mode while the additional branch collapses.  This identifies the extra
mode as the collective response of the magnetic carrier.

The longitudinal sector exposes the two-fluid nature of the system more
directly.  Without entrainment the electric and magnetic density
oscillations give two independent Langmuir modes.  With entrainment, the
off-diagonal inertial response mixes the two oscillations, turns their
crossing into an avoided crossing, and yields the stability condition
\[
        \kappa^2<1 ,
\]
equivalent to positive definiteness of the two-fluid momentum matrix.

We also examine a magnetically neutral branch obtained by resolving the
magnetic component into monopole and antimonopole species.  A magnetic
charge-conjugation symmetry \(\mathcal C_m\) selects a background with
vanishing net magnetic current, so the long-range monopole field is absent
even though the total magnetic density is nonzero.  The remaining
magnetic-charge fluctuations are screened, but they still leave a
collective response controlled by the magnetic plasma frequency.  Thus the
most robust observable signature is the density scaling
\[
        \omega_{\rm coll}^2\sim \omega_{pm}^2\propto n^0_{(m)} ,
\]
which may survive dissipative broadening in effective monopole media.

We also briefly comment on a possible structural implication of the same
\(\mathcal C_m\)-symmetric viewpoint.  A pair of dyons with opposite magnetic
charges can be magnetically neutral as a whole while carrying a nonzero
electric charge.  The mutual Dirac--Schwinger--Zwanziger pairing of the two
constituents can then constrain the residual electric charge even though no
free monopole appears in the asymptotic spectrum.  This observation is not used
in the linear plasma calculation and should be regarded as a possible
interpretive extension rather than as an input to the mode analysis.

The paper is organized as follows.  In Sec.~\ref{sec:modes} we derive the
linear transverse and longitudinal modes of the dual plasma in the absence of
entrainment.  In Sec.~\ref{sec:entrainment} we include entrainment, derive the
longitudinal avoided crossing and stability bound, analyze the
\(\mathcal C_m\)-protected neutral magnetic branch, and discuss screening and
thermal de-screening.  Section~\ref{sec:discussion} summarizes the physical
interpretation: static hiding of net magnetic charge, dynamical visibility
through collective response, and possible extensions to effective monopole
media, cosmology, and dyonic composites.

\section{Electromagnetic modes of the dual plasma}
\label{sec:modes}
We first examine the linear electromagnetic response of the dual plasma before
including entrainment. This already shows that the medium is not merely a
reparametrization of conventional monopole magnetohydrodynamics. The magnetic
fluid supplies an independent dynamical response to the magnetic field, in the
same sense that the electric fluid supplies the usual plasma response to the
electric field. As a result, the electromagnetic spectrum contains two plasma
scales and two propagating transverse branches.

We linearize the equations about a static, homogeneous, charge-neutral, and
unmagnetized background,
\[
\mathbf E_0=\mathbf B_0=0,\qquad
n_{(e)}=n_{(e)}^0,\qquad
n_{(m)}=n_{(m)}^0,
\]
with both fluids initially at rest. Perturbations are taken to be plane waves
\(\propto \exp[i(\mathbf k\cdot\mathbf x-\omega t)]\). Throughout this section
we work in the cold limit and neglect direct entrainment; the entrainment-induced
longitudinal mixing is treated in Sec.~\ref{sec:entrainment}.

\subsection{Setup: the closed two-fluid model}
\label{sec:setup}

We take as our starting point an effective closed two-fluid model with
separately conserved electric and magnetic charge currents.  The model is
defined by the duality-symmetric Maxwell equations
\begin{equation}
\nabla_b F^{ab}=j^a_{(e)},\qquad \nabla_b{}^{*}F^{ab}=j^a_{(m)} ,
\label{eq:setup-maxwell}
\end{equation}
together with the electric and magnetic Lorentz-force laws for the two
charge-carrying fluids.  The equations may be motivated by a convective
matter-space variational construction, but the linear analysis below uses
only the effective equations displayed in this section.
Electric and magnetic charge are carried by two matter-space flows with identically conserved number currents, $\nabla_a n^a_{(I)}=0$ ($I=e,m$), and fixed charge per particle $q,g$,
so that
\begin{equation}
j^a_{(e)}=q\,n^a_{(e)},\qquad j^a_{(m)}=g\,n^a_{(m)} .
\end{equation}
The two fluids satisfy the electric and magnetic Lorentz-force laws
\begin{equation}
n^b_{(e)}\bigl(\d\mu^{(e)}\bigr)_{ba}=q\,n^b_{(e)}F_{ba},\qquad
n^b_{(m)}\bigl(\d\mu^{(m)}\bigr)_{ba}=g\,n^b_{(m)}{}^{*}F_{ba} .
\label{eq:setup-lorentz}
\end{equation}
The inertia and the inter-fluid entrainment follow from a Carter master function
\begin{equation}
\Lambda=-\rho_e(n_e)-\rho_m(n_m)-\frac{\alpha}{2}x^2,
\qquad n_I=\sqrt{-n^a_{(I)}n_{(I)a}},\quad x^2=-n^a_{(e)}n_{(m)a},
\label{eq:setup-master}
\end{equation}
whose momenta $\mu^{(I)}_a=\partial\Lambda/\partial n^a_{(I)}$ take the entrained
form
\begin{equation}
\mu^{(I)}_a=\mathcal B^{(I)}n^{(I)}_a+\mathcal A\,n^{(J)}_a,
\qquad
\mathcal B^{(I)}=\frac{1}{n_I}\frac{\d\rho_I}{\d n_I},\quad
\mathcal A=\frac{\alpha}{2},\quad J\neq I .
\label{eq:setup-momenta}
\end{equation}
The off-diagonal coefficient $\mathcal A$ is the inertial entrainment between the
electric and magnetic fluids: upon linearization it supplies the off-diagonal
entry $\rho_{em}$ of the momentum matrix $M$ introduced below, and hence the
mixing parameter $\kappa^2=\rho_{em}^2/(\rho_{ee}\rho_{mm})$, while $\alpha=0$
decouples the fluids except through the field and gravity. When monopoles and
antimonopoles are resolved separately, the same framework yields an
exchange-symmetric pair whose intra-magnetic coupling $\beta$ is likewise an
ordinary Carter entrainment coefficient rather than an additional structure.

\subsection{Linear response}
The linearized Lorentz-force equations are
\begin{equation}
m_{(e)}\partial_t\mathbf v_{(e)}=q\,\mathbf E,
\qquad
m_{(m)}\partial_t\mathbf v_{(m)}=g\,\mathbf B .
\end{equation}
Thus the electric fluid is accelerated by the electric field, while the magnetic
fluid is accelerated by the magnetic field. For harmonic perturbations this
gives
\begin{equation}
\mathbf v_{(e)}=\frac{i q}{m_{(e)}\omega}\mathbf E,
\qquad
\mathbf v_{(m)}=\frac{i g}{m_{(m)}\omega}\mathbf B ,
\qarrow 
\mathbf j_{(e)}
=\frac{i\omega_{pe}^2}{\omega}\mathbf E,
\qquad
\mathbf j_{(m)}
=\frac{i\omega_{pm}^2}{\omega}\mathbf B ,
\label{eq:jEB}
\end{equation}
where the electric and magnetic plasma frequencies,
\begin{equation}
\omega_{pe}^2=\frac{q^2 n_{(e)}^0}{m_{(e)}},
\qquad
\omega_{pm}^2=\frac{g^2 n_{(m)}^0}{m_{(m)}} .
\label{eq}
\end{equation}
In a relativistic fluid these masses are replaced by the appropriate inertia per
particle,
\[
m_{(I)}\rightarrow h_I:=\frac{\rho_I+p_I}{n_{(I)}} .
\]
The two plasma frequencies are exchanged by electric--magnetic duality,
\[
q\leftrightarrow g,\qquad
n_{(e)}^0\leftrightarrow n_{(m)}^0,\qquad
\omega_{pe}\leftrightarrow\omega_{pm}.
\]

\subsection{Transverse modes}
\label{subsec:transverse-modes}

We now consider transverse electromagnetic waves.  
Take the wave vector along the \(z\)-axis and choose a transverse polarization,
$
\mathbf k=k\hat z,~
\mathbf E=E\,\hat x, ~
\mathbf B=B\,\hat y .
$
For perturbations proportional to
\(\exp[i(\mathbf k\cdot\mathbf x-\omega t)]\), the linearized Maxwell equations
are
\be{LinearMax}
\nabla\times\mathbf B-\partial_t\mathbf E = \mathbf j_{(e)},\qquad
\nabla\times\mathbf E+\partial_t\mathbf B = -\mathbf j_{(m)} .
\ee
Using the linear response currents~\eqref{eq:jEB}, these equations become
\be{lineareq1}
\left(\omega-\frac{\omega_{pe}^2}{\omega}\right)E
=-kB, \qquad
kE
=-\left(\omega-\frac{\omega_{pm}^2}{\omega}\right)B .
\ee 
The relative signs depend on the polarization convention, but the dispersion
relation does not.  Eliminating \(E\) and \(B\) gives
\begin{equation}
\boxed{
\left(\omega^2-\omega_{pe}^2\right)
\left(\omega^2-\omega_{pm}^2\right)
=\omega^2 k^2
} \qarrow 
\omega^4
-\left(k^2+\omega_{pe}^2+\omega_{pm}^2\right)\omega^2
+\omega_{pe}^2\omega_{pm}^2
=0 .
\label{eq:transverse-quartic}
\end{equation}
Thus the two transverse branches are
\begin{equation}
\omega_\pm^2= \frac{1}{2}
\left[
k^2+\omega_{pe}^2+\omega_{pm}^2
\pm
\sqrt{
\left(k^2+\omega_{pe}^2+\omega_{pm}^2\right)^2
-4\omega_{pe}^2\omega_{pm}^2
}
\right].
\label{eq:transverse-branches}
\end{equation}
This expression is manifestly invariant under electric--magnetic duality,
$
\omega_{pe}\leftrightarrow\omega_{pm}.
$
\begin{figure}[htbp]
\centering
\begin{subfigure}{0.55\textwidth}
    \includegraphics[width=\linewidth]{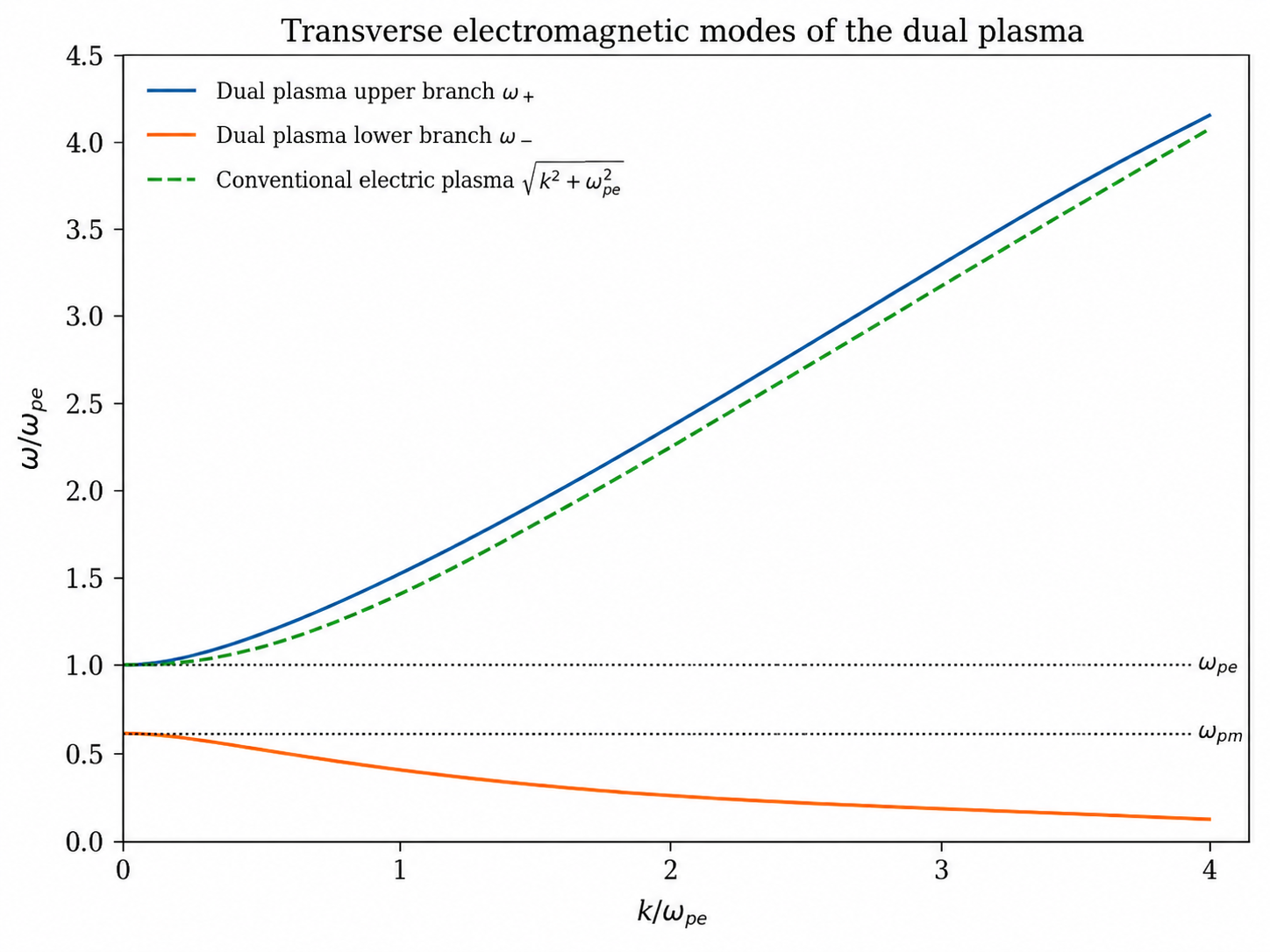}
\end{subfigure}
\caption{Transverse electromagnetic spectrum of the dual plasma.  The electric and magnetic plasma frequencies generate two duality-related cutoffs and two propagating transverse branches. 
}
\label{fig1}
\end{figure}
The two transverse branches in Eq.~\eqref{eq:transverse-branches} are shown in Fig.~\ref{fig1}.
At zero wave number the two cutoffs are
\begin{equation}
\omega_+^2(0)=\max(\omega_{pe}^2,\omega_{pm}^2),
\qquad
\omega_-^2(0)=\min(\omega_{pe}^2,\omega_{pm}^2).
\end{equation}
Thus the transverse response contains two plasma scales rather than one.  If the
magnetic plasma frequency is taken to zero, one recovers the usual electric
plasma branch together with a collapsed zero-frequency branch,
\begin{equation}
\omega_+^2=k^2+\omega_{pe}^2,
\qquad
\omega_-^2=0
\qquad
(\omega_{pm}=0).
\end{equation}
The extra branch is therefore not a relabeling of the ordinary electric plasma
mode; it is the transverse dynamical response of the magnetic fluid.

The transverse sector is linearly stable.  The discriminant satisfies
\begin{equation}
\left(k^2+\omega_{pe}^2+\omega_{pm}^2\right)^2
-4\omega_{pe}^2\omega_{pm}^2
=
k^4+2k^2(\omega_{pe}^2+\omega_{pm}^2)
+(\omega_{pe}^2-\omega_{pm}^2)^2
\ge 0 ,
\end{equation}
and both roots in Eq.~\eqref{eq:transverse-branches} are non-negative.  The
upper branch approaches the vacuum light cone at large \(k\),
\begin{equation}
\omega_+^2=k^2+\omega_{pe}^2+\omega_{pm}^2
+O(k^{-2}),
\end{equation}
whereas the lower branch becomes
\begin{equation}
\omega_-^2=\frac{\omega_{pe}^2\omega_{pm}^2}{k^2}
+O(k^{-4}) .
\end{equation}
The lower branch is therefore a medium-supported transverse mode.  Its presence,
together with the duality-related pair of cutoffs, is one of the simplest
linear signatures that the dual plasma is physically distinct from conventional
monopole magnetohydrodynamics.

\subsection{Longitudinal modes without entrainment}
In the absence of entrainment, the longitudinal electric and magnetic density
oscillations decouple. For a longitudinal electric/magnetic perturbation, Gauss' law and charge conservation give the usual Langmuir mode,
\begin{equation}
\omega^2=\omega_{pe}^2 , \qquad 
\omega^2=\omega_{pm}^2 .
\end{equation}
Thus the cold, non-entrained dual plasma contains two independent longitudinal
plasma oscillations. This result should not be confused with conventional
monopole magnetohydrodynamics, where the magnetic charge current is typically
introduced phenomenologically into Maxwell's equations. Here the magnetic
current is carried by an independent matter-space fluid and therefore has its
own inertia, its own plasma frequency, and its own density mode.

The decoupling in this section is not protected once the two fluids are
entrained. Since entrainment couples the momenta of the electric and magnetic
constituents, it mixes precisely the longitudinal density oscillations described
above. The resulting avoided crossing and stability boundary are derived in
Sec.~\ref{sec:entrainment}.

\subsection{Physical interpretation}
The mode spectrum gives the first sense in which hidden magnetic charge can be dynamically visible. 
A neutral and unmagnetized background has no static Coulomb-like magnetic field, and therefore no direct monopole signal in the background electromagnetic field. 
Nevertheless, the magnetic fluid is not dynamically inert. 
Its inertia and charge enter the linear response through \(\omega_{pm}\), producing a second transverse cutoff, a second transverse branch, and a magnetic Langmuir mode.

This distinction will be important below. 
The absence of a static monopole field in a neutral background does not imply the absence of observable magnetic matter. 
Rather, the magnetic sector is probed through collective response. 
In the symmetric monopole--antimonopole plasma discussed in Sec.~\ref{subsec:Cm-protection}, this statement becomes sharper: net magnetic charge can be protected and strongly screened, while the hidden sector still leaves dynamical signatures in the wave spectrum and in the longitudinal electric--magnetic mixing.

\section{Entrainment, protected neutrality, and dynamical visibility}
\label{sec:entrainment}
The transverse spectrum of Sec.~\ref{sec:modes} already shows that a dual plasma
has more structure than a conventional monopole magnetohydrodynamic medium.  We
now turn to the part of the response where the two-fluid nature is most direct:
the longitudinal sector.  In a longitudinal perturbation both fluids move along
the same wave vector, so any inertial coupling between the constituents enters
without reference to polarization.  This is the cleanest setting in which to
isolate the effect of matter-space entrainment.

We parameterize the cold linearized momenta by
\begin{equation}
\mathbf p_{(I)}=
\sum_J \mathsf M_{IJ}\mathbf v_{(J)},
\qquad
\mathsf M
=\begin{pmatrix}
\rho_{ee} & \rho_{em} \\
\rho_{em} & \rho_{mm}
\end{pmatrix},
\qquad
\kappa^2
:=
\frac{\rho_{em}^2}{\rho_{ee}\rho_{mm}} .
\label{eq:momentum-matrix}
\end{equation}
Here \(\rho_{ee}\) and \(\rho_{mm}\) are the electric and magnetic inertial
densities, while \(\rho_{em}\) is the entrainment coefficient.  In a cold
nonrelativistic limit one may identify
\[
\rho_{ee}\simeq m_{(e)}n_{(e)}^0,
\qquad
\rho_{mm}\simeq m_{(m)}n_{(m)}^0,
\]
whereas in a relativistic fluid the masses are replaced by the appropriate
enthalpy per particle.  The parameter \(\kappa\) measures the strength of
off-diagonal inertial mixing.  Absence of ghosts in the kinetic energy requires
\begin{equation}
\det \mathsf M>0
\qquad\Longleftrightarrow\qquad
\kappa^2<1 .
\label{eq:ghost-free-kappa}
\end{equation}

In this section we focus on the longitudinal entrainment problem.  The transverse
branches derived in Sec.~\ref{subsec:transverse-modes} are therefore used as the
unentrained electromagnetic reference spectrum.  A fully transverse entrained
response would require keeping the vector off-diagonal momentum response for
both polarizations.  That effect is not needed for the longitudinal stability
and neutrality mechanism developed below.

\subsection{Longitudinal mixing}
\label{subsec:longitudinal-mixing}

Take a longitudinal plane wave with
$
\mathbf k=k\hat z,
~~
\mathbf v_{(e)}=v_{(e)}\hat z,
~~
\mathbf v_{(m)}=v_{(m)}\hat z .
$
The cold longitudinal force equations are
\begin{equation}
-i\omega
\begin{pmatrix}
p_{(e)}\\
p_{(m)}
\end{pmatrix}
=
\begin{pmatrix}
q n^0_{(e)} E_L\\
g n^0_{(m)} B_L
\end{pmatrix},
\qquad
\begin{pmatrix}
p_{(e)}\\
p_{(m)}
\end{pmatrix}
=
\begin{pmatrix}
\rho_{ee} & \rho_{em}\\
\rho_{em} & \rho_{mm}
\end{pmatrix}
\begin{pmatrix}
v_{(e)}\\
v_{(m)}
\end{pmatrix}.
\label{eq:longitudinal-force-matrix}
\end{equation}
The longitudinal fields are fixed by the electric and magnetic Gauss laws.  For
harmonic perturbations, charge conservation gives
\[
\delta n_{(e)}=\frac{k n_{(e)}^0}{\omega}v_{(e)},
\qquad
\delta n_{(m)}=\frac{k n_{(m)}^0}{\omega}v_{(m)} ,
\]
and hence the restoring forces are controlled by the two plasma frequencies
\begin{equation}
\omega_{pe}^2=\frac{q^2 \big(n_{(e)}^0\big)^2}{\rho_{ee}},
\qquad
\omega_{pm}^2=\frac{g^2 \big(n_{(m)}^0\big)^2}{\rho_{mm}} .
\label{eq:entrained-plasma-frequencies}
\end{equation}
Equivalently, after dividing by the diagonal inertias, the longitudinal
eigenvalue problem may be written as
\begin{equation}
\begin{pmatrix}
\omega^2-\omega_{pe}^2 & \kappa\,\omega^2\\
\kappa\,\omega^2 & \omega^2-\omega_{pm}^2
\end{pmatrix}
\begin{pmatrix}
v_{(e)}\\
v_{(m)}
\end{pmatrix}
=0 ,
\label{eq:longitudinal-eigenproblem}
\end{equation}
where the sign of \(\kappa\) can be absorbed into the relative phase of the two
velocity perturbations.  The determinant condition gives
\begin{equation}
\boxed{\;
(1-\kappa^2)\omega^4
-(\omega_{pe}^2+\omega_{pm}^2)\omega^2
+\omega_{pe}^2\omega_{pm}^2
=0
\;}
\label{eq:disp-L}
\end{equation}
This reduces to the two independent Langmuir oscillations when \(\kappa=0\).

\begin{figure}[htbp]
\centering
\begin{subfigure}{0.55\textwidth}
    \centering
    \includegraphics[width=\linewidth]{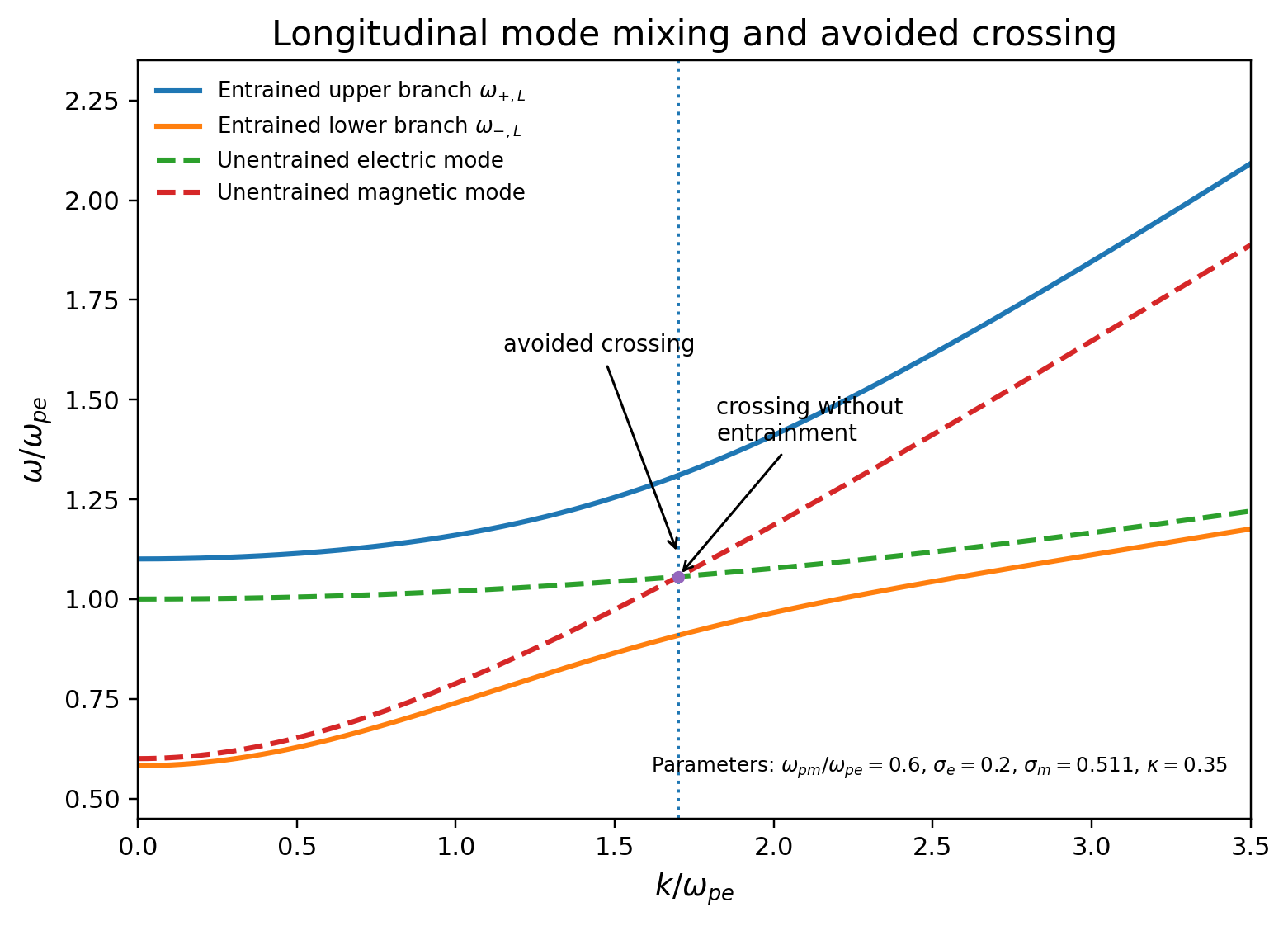}
\end{subfigure}
\caption{Longitudinal electric--magnetic mode mixing in the presence of thermal pressure.  Without entrainment the electric and magnetic plasma oscillations cross.  Entrainment turns the crossing into an avoided crossing.
}
\label{fig1b}
\end{figure}

The two longitudinal eigenfrequencies are
\begin{equation}
\omega_{\pm,L}^2=
\frac{
\omega_{pe}^2+\omega_{pm}^2
\pm
\sqrt{
(\omega_{pe}^2-\omega_{pm}^2)^2
+4\kappa^2\omega_{pe}^2\omega_{pm}^2
}
}
{2(1-\kappa^2)} .
\label{eq:longitudinal-branches}
\end{equation}
Thus entrainment does not merely shift the plasma frequencies.  It mixes the electric and magnetic density oscillations as shown in the $k=0$ values in Fig.~\ref{fig1b}.
Near resonance, \(\omega_{pe}\simeq\omega_{pm}\), the eigenvectors are approximately the
in-phase and out-of-phase combinations of the two fluids, and the frequency
crossing is avoided.  At exact resonance,
\(\omega_{pe}=\omega_{pm}=\omega_p\), one finds
\begin{equation}
\omega_{\pm,L}^2
=
\frac{\omega_p^2}{1\mp\kappa}.
\label{eq:resonant-longitudinal-splitting}
\end{equation}
The splitting is therefore controlled directly by the entrainment strength.

Thermal pressure or finite compressibility dresses the two diagonal restoring
forces.  To leading order one may write
\begin{equation}
\omega_{pI}^2
\longrightarrow
\omega_{pI}^2+\sigma_I^2 k^2 ,
\qquad
I=e,m ,
\label{eq:thermal-dressed-plasma-frequency}
\end{equation}
where \(\sigma_I\) is the sound speed of species \(I\).  
As shown in Fig.~\ref{fig1b}, this moves the location
of the avoided crossing but does not change the ghost-free condition \(\kappa^2<1\).  
If \(\sigma_{(e)}\neq\sigma_{(m)}\), the crossing occurs near
\begin{equation}
k_*^2=
\frac{\omega_{pe}^2-\omega_{pm}^2}
{\sigma_{(m)}^2-\sigma_{(e)}^2}
\label{eq:avoided-crossing-kstar}
\end{equation}
when the right-hand side is positive.

\subsection{Stability}
\label{subsec:longitudinal-stability}

\begin{figure}[htbp]
\centering
\begin{subfigure}{0.55\textwidth}
    \includegraphics[width=\linewidth]{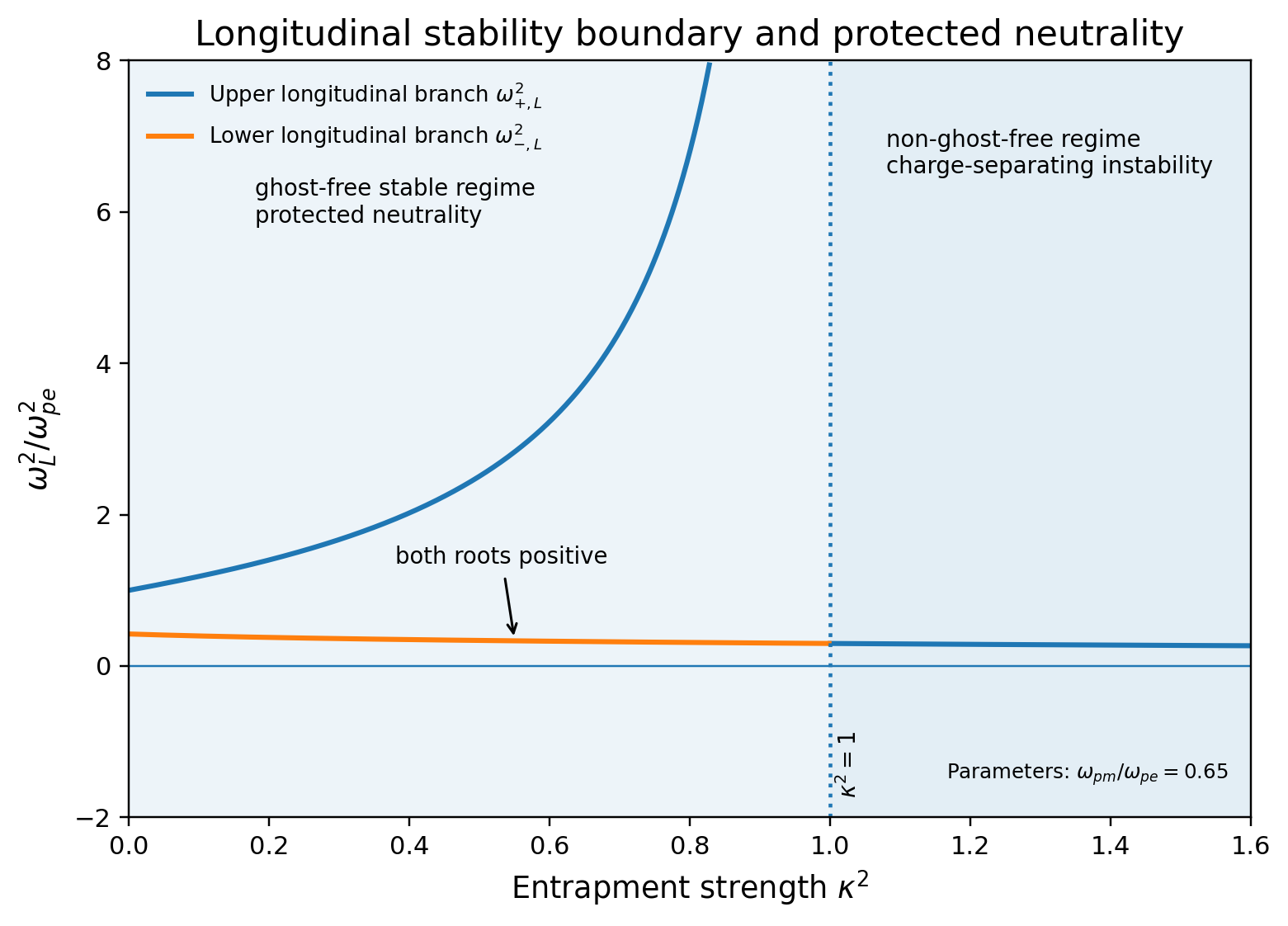}
\end{subfigure}
\caption{Longitudinal stability boundary.  In the ghost-free regime
\(\kappa^2<1\), both longitudinal modes have positive \(\omega^2\).  The
charge-separating instability appears only when the inertial matrix loses
positive definiteness at \(\kappa^2=1\).
}
\label{fig2}
\end{figure}

The stability condition follows directly from Eq.~\eqref{eq:disp-L}.  The
product of the two roots is
\begin{equation}
\omega_{+,L}^2\omega_{-,L}^2
=\frac{\omega_{pe}^2\omega_{pm}^2}{1-\kappa^2}.
\end{equation}
For \(\kappa^2<1\) this product is positive, and the sum of the two roots is also
positive,
\begin{equation}
\omega_{+,L}^2+\omega_{-,L}^2
=
\frac{\omega_{pe}^2+\omega_{pm}^2}{1-\kappa^2}>0 .
\end{equation}
Thus both longitudinal modes are stable in the ghost-free regime.  For
\(\kappa^2>1\), the product of the roots is negative, so one mode has
\(\omega^2<0\).  The onset of the longitudinal instability therefore coincides
with the loss of positive definiteness of the inertial matrix:
\begin{equation}
\boxed{\;
\text{longitudinal stability}
\quad\Longleftrightarrow\quad
\mathsf M>0
\quad\Longleftrightarrow\quad
\kappa^2<1 .
\;}
\label{eq:stability-positive-matrix}
\end{equation}
The schematic picture for this stability is given in Fig.~\ref{fig2}.

This is a genuine two-fluid effect.  The electric and magnetic fluids do not
exchange charge, and the mixing is not generated by a direct conversion process.
It is inertial: the momentum of each constituent contains a component
proportional to the velocity of the other.  Consequently the electric and
magnetic Langmuir modes repel each other in the spectrum.  This avoided crossing
and its associated stability boundary are absent from conventional monopole MHD,
where magnetic charge currents are introduced phenomenologically rather than as
independent matter-space fluids with their own momenta.

\subsection{Magnetic charge conjugation and protected neutrality}
\label{subsec:Cm-protection}

We now refine the magnetic sector by resolving it into monopole and antimonopole
fluids.  Let the two magnetic species be denoted \(m+\) and \(m-\), with
independent matter spaces \(\mathcal M_{(m+)}\) and \(\mathcal M_{(m-)}\).  Their
number-current three-forms are \(N^{(m+)}\) and \(N^{(m-)}\), with dual current
vectors \(n^a_{(m+)}\) and \(n^a_{(m-)}\).  The two species carry opposite
matter-space non-closures\footnote{We barrow this form from the matter space framework. The two matter spaces ($m\pm$) carry opposite magnetic charges.},
\begin{equation}
\d_{\mathcal M}G^{(m+)}=
g\,\Omega^{(m+)},
\qquad
\d_{\mathcal M}G^{(m-)}=
-g\,\Omega^{(m-)} .
\label{eq:opposite-magnetic-nonclosure}
\end{equation}
After pull-back,
\begin{equation}
J^{(m+)}=\d G^{(m+)}=gN^{(m+)},
\qquad
J^{(m-)}=\d G^{(m-)}=-gN^{(m-)} .
\label{eq:opposite-magnetic-currents}
\end{equation}
The net magnetic current three-form and its dual vector are therefore
\begin{equation}
J^{(m)}_{\rm net}=g\big(N^{(m+)}-N^{(m-)}\big),
\qquad
j^a_{(m),{\rm net}}=g\big(n^a_{(m+)}-n^a_{(m-)}\big).
\label{eq:net-magnetic-current}
\end{equation}

Now, we define magnetic charge conjugation by exchanging the two species together with their non-closure data,
\begin{equation}
\mathcal C_m:\qquad
M^A_{(m+)}\leftrightarrow M^A_{(m-)},
\qquad
G^{(m+)}\leftrightarrow G^{(m-)} .
\label{eq:Cm-action}
\end{equation}
Under this transformation, the net magnetic charge change sign:
\[
N^{(m+)}\leftrightarrow N^{(m-)},
\qquad
J^{(m)}_{\rm net}\mapsto -J^{(m)}_{\rm net}.
\]
A \(\mathcal C_m\)-invariant background must therefore obey
\begin{equation}
N^{(m+)}=N^{(m-)},
\qquad
j^a_{(m),{\rm net}}=0 .
\label{eq:Cm-neutral-background}
\end{equation}
Net magnetic neutrality is then not an assumption imposed after the fact.  It is
the symmetry condition selecting the \(\mathcal C_m\)-even branch of the magnetic
fluid.

To see whether this neutral branch is dynamically stable, consider a
\(\mathcal C_m\)-invariant master function for the two magnetic species,
\begin{equation}
\Lambda_m=
-\rho_m(n_{(m+)})
-\rho_m(n_{(m-)})
-\frac{\beta}{2}y^2 ,
\qquad
y^2=-n^a_{(m+)}n_{(m-)a}, \qquad \beta \geq 0.
\label{eq:magnetic-pair-master}
\end{equation}
For a static homogeneous background, the magnetic contribution to the energy
density is
\begin{equation}
E_m(n_{(m+)},n_{(m-)})=\rho_m(n_{(m+)})
+\rho_m(n_{(m-)})
+\frac{\beta}{2}n_{(m+)}n_{(m-)} .
\label{eq:magnetic-pair-energy}
\end{equation}
At fixed total magnetic density
\[
N=n_{(m+)}+n_{(m-)},
\qquad
n_{(m\pm)}=\frac{N}{2}\pm\delta ,
\]
one obtains
\begin{equation}
E_m=
2\rho_m(N/2)
+\frac{\beta N^2}{8}
+
\left[
\rho_m''(N/2)-\frac{\beta}{2}
\right]\delta^2
+O(\delta^4).
\label{eq:Cm-energy-expansion}
\end{equation}
The charge-conjugation symmetric point is therefore a local minimum if and only if
\begin{equation}
\boxed{\;
2\rho_m''(N/2)-\beta>0 .
\;}
\label{eq:Cm-threshold}
\end{equation}

The same condition is the ghost-free condition for the two magnetic species.
Indeed, the Hessian of \(E_m\) at the symmetric point is
\begin{equation}
H
=
\begin{pmatrix}
\rho_m'' & \beta/2\\
\beta/2 & \rho_m''
\end{pmatrix},
\qquad
\lambda_{\pm}=
\rho_m''\pm\frac{\beta}{2}.
\label{eq:magnetic-pair-hessian}
\end{equation}
The \(\mathcal C_m\)-breaking direction is the antisymmetric fluctuation
\((1,-1)\), whose eigenvalue is
\[
\lambda_-=\rho_m''-\frac{\beta}{2}.
\]
Thus the neutral branch is stable precisely when the antisymmetric mode is not a
ghost.  In the notation of Eq.~\eqref{eq:momentum-matrix}, this may be written as
\begin{equation}
\kappa_{mm}^2
:=
\frac{(\beta/2)^2}{(\rho_m'')^2}
<1 .
\label{eq:kappa-mm}
\end{equation}
A spontaneous transition to a state with
\(J^{(m)}_{\rm net}\neq0\) can occur only when this bound is violated.

\begin{proposition}[Dynamical protection of magnetic neutrality]
\label{prop:Cm-protection}
In the \(\mathcal C_m\)-invariant monopole--antimonopole fluid, the neutral
background \(N^{(m+)}=N^{(m-)}\) is energetically stable throughout the
ghost-free regime.  A charge-separating instability, and hence a nonzero net
magnetic current \(J^{(m)}_{\rm net}\), appears only when the antisymmetric
longitudinal mode has already crossed the stability boundary.
\end{proposition}

This is the precise sense in which net magnetic neutrality is protected.  The
theory does not require one to impose a vanishing monopole density by hand.
Rather, the \(\mathcal C_m\)-even state  is dynamically selected throughout the regime in which the effective two-fluid description has positive kinetic energy.

\subsection{Screening, strong coupling, and thermal de-screening}
\label{subsec:screening-descreening}

Protected neutrality closes the direct static channel for observing a monopole
plasma.  In the \( \mathcal C_m \)-symmetric branch, the long-range \(1/r^2\) field
of the net magnetic charge cancels because
\[
J^{(m)}_{\rm net}=0 .
\]
This cancellation, however, should not be confused with the absence of magnetic
matter.  The individual \(m+\) and \(m-\) fluids still fluctuate, and their
fluctuations are screened over the magnetic Debye length.

\begin{figure}[htbp]
\centering
\begin{subfigure}{0.55\textwidth}
    \centering
    \includegraphics[width=\linewidth]{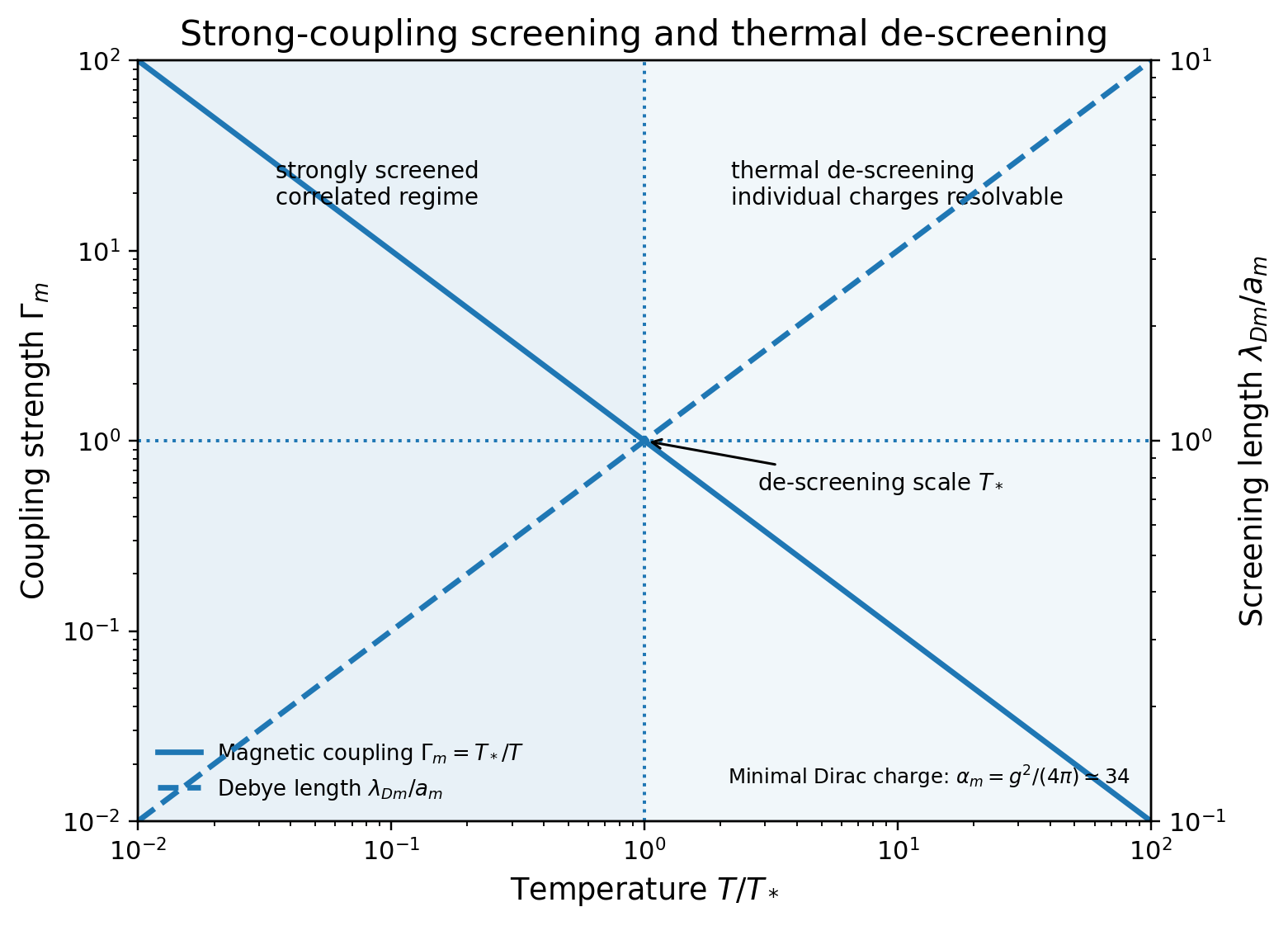}
\end{subfigure}
\caption{Strong-coupling screening and thermal de-screening.  Because the magnetic coupling is large, the monopole plasma remains strongly screened for \(T<T_*\).  Individual magnetic charges become thermally resolvable only above the de-screening scale.
}
\label{fig2b}
\end{figure}

For a magnetic plasma with total magnetic density \(n_{(m)}^0\), the magnetic plasma frequency is
\begin{equation}
\omega_{pm}^2=\frac{g^2 n_{(m)}^0}{m_{(m)}} , \qquad
n_{(m)}^0=n_{(m+)}^0+n_{(m-)}^0 , 
\label{eq:magnetic-plasma-frequency-total}
\end{equation}
where the contributions of the two species add because screening depends on
\(g^2\), not on the sign of \(g\).  In a classical thermal regime, the magnetic Debye length
\begin{equation}
\lambda_{Dm}^2=\frac{v_{th,m}^2}{\omega_{pm}^2}=\frac{T/m_{(m)}}{g^2n_{(m)}^0/m_{(m)}}
=\frac{T}{g^2n_{(m)}^0}\propto T ,
\label{eq:magnetic-debye-length}
\end{equation}
lengthens with temperature.
Whether this screening still hides individual
charges is controlled by the magnetic plasma coupling at the Wigner--Seitz
radius $a_m=(n_{(m)}^0)^{-1/3}$,
\begin{equation}
\Gamma_m=\frac{g^2/(4\pi a_m)}{T}=\frac{\alpha_m\,(n_{(m)}^0)^{1/3}}{T},
\qquad
N_D=n_{(m)}^0\lambda_{Dm}^3=\frac{T^{3/2}}{g^3 (n_{(m)}^0)^{1/2}}
=\left(4\pi \Gamma_m\right)^{-3/2},
\end{equation}
as shown in Fig.~\ref{fig2b}, Dirac quantization gives a strong magnetic coupling,
\begin{equation}
\alpha_m= \frac{g^2}{4\pi} =
\frac{1}{4\alpha}
\simeq 34 ,
\label{eq:large-magnetic-coupling}
\end{equation}
for the minimal Dirac charge in the convention \(qg=2\pi\).
So neutrality and screening persist up to a temperature about 34 times the naive ``Coulomb energy $\sim$ thermal energy'' estimate: the strong coupling that enhances screening in Sec.~\ref{subsec:Cm-protection} also delays de-screening. 
For $\Gamma_m\gg1$ ($N_D\ll1$, low $T$) the medium is a strongly coupled, correlated neutral plasma: $\lambda_{Dm}\ll a_m$ and each monopole field is screened well inside the interparticle spacing---invisible.
Because \(\alpha_m\) is large, residual magnetic-charge fluctuations are strongly screened.  

Individual magnetic charges become thermally resolvable only when the
temperature is high enough that the magnetic Debye length is comparable to, or larger than, the interparticle spacing. 
For $\Gamma_m\lesssim1$ ($N_D\gtrsim1$, high
$T$) the plasma is weakly coupled: $\lambda_{Dm}\gtrsim a_m$, and over the window
$a_m\lesssim r\lesssim\lambda_{Dm}$ the bare $g/r^2$ field of an individual
charge is resolved. 
This gives the de-screening scale
\begin{equation}
\boxed{\;
T_*
\sim
\alpha_m (n_{(m)}^0)^{1/3}=
\frac{g^2}{4\pi}(n_{(m)}^0)^{1/3}.
\;}
\label{eq:Tstar}
\end{equation}
The large value of \(\alpha_m\) therefore delays thermal de-screening:
the static monopole channel remains hidden unless the temperature is
parametrically high compared with the naive density scale
\((n_{(m)}^0)^{1/3}\).

Crossing \(T_*\) does not generate a net magnetic charge.  The condition
\(J^{(m)}_{\rm net}=0\) remains enforced by \(\mathcal C_m\) as long as the
symmetry is unbroken.  What changes is the visibility of the individual
\(\pm g\) constituents.  At low temperature, the plasma is strongly coupled and
correlated, and the field of each charge is screened inside the interparticle
scale.  At high temperature, screening weakens and individual charges can be
resolved over distances between \(a_m\) and \(\lambda_{Dm}\).

\subsection{Interpretation}
\label{subsec:dynamic-visibility-interpretation}

The result is not that magnetic charge is simply unobservable.  Rather, the
static monopole signal is removed in the \(\mathcal C_m\)-symmetric branch, while
the magnetic sector remains visible through collective response.  The transverse
spectrum contains a magnetic plasma scale, the longitudinal sector exhibits
electric--magnetic mode mixing, and entrainment produces an avoided crossing and
a stability boundary.  These are dynamical signatures of the hidden magnetic
fluid.

Thus the dual plasma realizes a separation between two notions of observability.
As an isolated static source, net magnetic charge is screened or symmetry
forbidden in the stable neutral branch.  As a dynamical medium, however, the
magnetic sector leaves measurable imprints in the electromagnetic response.  The
appropriate conclusion is therefore not absolute invisibility, but static hiding
together with dynamical visibility.

\subsection{Experimental observability}
The ideal transverse cutoffs and sharp propagating branches derived above should
not be taken as the most realistic experimental targets in ordinary spin ice.
In spin-ice materials, where magnetic monopole excitations are realized
effectively~\cite{CastelnovoMoessnerSondhi}, monopole motion is strongly
dissipative, and the clean collisionless dual-plasma dispersion is likely to be
broadened into a relaxation spectrum.  
A more robust signature is instead the density dependence of the collective magnetic-charge response.  Since the magnetic plasma scale is
\[
\omega_{pm}^2=\frac{g^2 n_{(m)}^0}{m_{(m)}} ,
\]
or, more generally, with \(m_{(m)}\) replaced by the appropriate effective
inertia, the characteristic magnetic-charge relaxation or resonance frequency
should scale with the monopole density as
\[
\omega_{pm}^2\propto n_{(m)}^0 .
\]
This scaling survives even when the ideal mode is damped, because it follows
from the restoring force associated with magnetic charge density fluctuations
rather than from the existence of an infinitely sharp quasiparticle pole.

Artificial spin ice provides a particularly promising arena for this test.  In
engineered two- or three-dimensional spin-ice architectures, the monopole density
can be tuned by temperature, field history, or sample preparation, and the
magnetic-charge configuration can be imaged directly.  Microwave spectroscopy,
Brillouin-light-scattering spectroscopy, or related dynamical probes could then
be used to correlate the measured relaxation or resonance scale with the
independently prepared monopole density.  The cleanest evidence for the dual
plasma response would not necessarily be a sharp transverse cutoff, but a
systematic scaling of the collective magnetic-charge spectral feature according
to
\[
\omega_{\rm coll}^2 \sim \omega_{pm}^2 \propto n_{(m)}^0 .
\]
Such an observation would distinguish a genuine magnetic-charge plasma response
from a purely phenomenological monopole current, and would provide an
experimentally accessible remnant of the dynamical visibility mechanism proposed
here.

If the early universe selects the \(\mathcal C_m\)-symmetric branch, monopoles
are produced with vanishing net magnetic charge but with a nonzero total
magnetic density.  In that case the relevant cosmological observable is not a
long-range monopole field, since \(J^{(m)}_{\rm net}=0\), but the collective
magnetic plasma scale
\[
\omega_{pm}^2(z)=\frac{g^2 n_{(m)}(z)}{m_{(m)}} .
\]
This scale can modify the propagation and damping of primordial electromagnetic
fields, and may leave imprints in the survival of primordial magnetic fields,
CMB-era plasma transfer, or hidden-sector radiation dynamics.  Thus a neutral
monopole plasma would be cosmologically visible, if at all, through its
density-dependent collective response rather than through a static monopole
Coulomb field.

\section{Discussion}
\label{sec:discussion}
 
We have studied the linear response of a dual plasma in which electric and
magnetic charges are carried by independent matter-space fluids.  The main
conclusion is that magnetic charge need not be directly visible as an isolated
static source in order to have observable consequences.  In the stable neutral
branch the long-range monopole field is absent or screened, yet the magnetic
sector remains visible both through the collective dynamics of the medium and
through the quantization of the electric charge it leaves behind.
 
There are three main dynamical results.  First, the transverse electromagnetic spectrum contains
two plasma scales: the electric and magnetic plasma frequencies enter
symmetrically, giving two duality-related cutoffs and two transverse branches.
The lower branch collapses to zero frequency when \(\omega_{pm}\to0\), showing
that it is a genuine response of the magnetic component rather than a
relabeling of the ordinary electric mode.
 
Second, the longitudinal sector exposes the two-fluid character of the theory.
The cold non-entrained limit gives two independent Langmuir modes; finite
compressibility dresses them by \(k^2\)-dependent corrections and allows them to
cross; matter-space entrainment then mixes the two density oscillations and
turns the crossing into an avoided crossing.  The same inertial mixing yields a
sharp stability condition \(\kappa^2<1\), equivalent to positive definiteness of
the momentum matrix.  The avoided crossing and the threshold are therefore
specific consequences of treating the magnetic carrier as an independent fluid
with its own momentum, not phenomenological additions to monopole
magnetohydrodynamics.
 
Third, resolving the magnetic component into monopole and antimonopole fluids
shows how net magnetic neutrality is dynamically protected.  In a
\(\mathcal C_m\)-symmetric sector the neutral branch satisfies
\(J^{(m)}_{\rm net}=0\); charge separation is the antisymmetric \(m+\!-m-\)
direction, and the onset of the charge-separating instability coincides with the
loss of the ghost-free condition.  The residual magnetic-charge fluctuations are
Debye screened, with a screening strength enhanced by the large magnetic
coupling \(\alpha_m=g^2/4\pi\simeq 1/4\alpha\).  This is an internal
effective-field-theory mechanism for the long-range static invisibility of a
neutral, strongly coupled monopole plasma, with a thermal de-screening scale
\(T_*\sim\alpha_m (n_{(m)}^0)^{1/3}\) above which the individual \(\pm g\)
constituents become resolvable while the net charge stays zero.

Taken together, the dual plasma realizes a separation between three notions of
observability.  Statically, the channel is closed: there is no unscreened
\(1/r^2\) monopole field in the stable \(\mathcal C_m\)-symmetric branch.
Dynamically, the channel is open: the magnetic plasma scale \(\omega_{pm}\),
longitudinal electric--magnetic mixing, entrainment-controlled stability, and
thermal de-screening are collective signatures of the hidden sector.
Structurally, the hidden dyonic substructure leaves a permanent imprint in the
quantization of electric charge, resolvable directly only above the binding
scale.  Magnetic charge is thus statically hidden, dynamically visible, and
structurally imprinted.
 
From an experimental point of view, the clean ideal-plasma transverse cutoffs may
not be the most robust target in ordinary spin ice, where monopole motion is
strongly dissipative.  A more realistic observable is the density dependence of a
collective magnetic-charge relaxation or resonance scale.  Since
\(\omega_{pm}^2=g^2 n_{(m)}^0/m_{(m)}\), or more generally with \(m_{(m)}\)
replaced by an effective inertia, the characteristic scale should obey
\(\omega_{\rm coll}^2\sim\omega_{pm}^2\propto n_{(m)}^0\).  This scaling survives
damping because it follows from the restoring force associated with
magnetic-charge density fluctuations rather than from a sharp quasiparticle pole.
Artificial spin ice, especially engineered two- or three-dimensional
architectures, offers a natural arena: the monopole density can be prepared and
imaged while microwave, Brillouin-light-scattering, or related probes test the
predicted scaling.
 
The construction treats magnetic charge as an effective matter-space
constituent.  It does not assume a GUT monopole, nor does it fix the inertial
mass: Dirac quantization constrains \(g\), not the inertia in the plasma
frequency, so the observable scale is \(\omega_{pm}^2=g^2 n_{(m)}^0/m_{\rm eff}\)
with \(m_{\rm eff}\) the appropriate inertial parameter.  
The usual cosmological monopole problem concerns the abundance of relic
magnetic charges~\cite{ZeldovichKhlopov,Preskill}; here the relevant question is instead whether a \(\mathcal C_m\)-symmetric population can be neutral in net magnetic charge while still carrying a collective plasma response.
If the early universe approximately respects
\(\mathcal C_m\), monopoles and antimonopoles are produced symmetrically, so the
symmetry suppresses the net asymmetry \(J^{(m)}_{\rm net}\propto N^{(m+)}-N^{(m-)}\)
rather than the total abundance \(N^{(m+)}+N^{(m-)}\).  A neutral monopole plasma
then produces no long-range cosmological monopole field, but its
density-dependent collective scale \(\omega_{pm}(z)\) could modify the
propagation or damping of primordial electromagnetic fields, or appear as a
hidden-sector plasma response.

 A further structural implication concerns electric charge quantization.
The matter-space construction does not require a constituent to be purely
electric or purely magnetic.  A pair of dyons with opposite magnetic
charges may form a magnetically neutral composite,
\[
(q,g)+(q',-g)=(Q_{\rm net},0),\qquad Q_{\rm net}=q+q' .
\]
Although such an object has no long-range monopole field, its internal
two-body structure is not equivalent to two purely electric charges.  The
Dirac--Schwinger--Zwanziger pairing of the two constituents is
\[
D_{12}=q(-g)-q'g=-Q_{\rm net}g .
\]
The possibility of constituents carrying both electric and magnetic charge is
standard in dyonic charge-lattice physics~\cite{Schwinger1966,Zwanziger1968,WittenEffect}.
Quantum consistency requires \(D_{12}\in 2\pi\mathbb Z\).  For the minimal
magnetic charge \(g=2\pi/e\), this gives
\[
Q_{\rm net}\in e\mathbb Z .
\]
Thus, if ordinary electric charges are interpreted as
\(\mathcal C_m\)-even dyonic composites, their net magnetic charge is
hidden while their surviving electric charge is quantized.  This
observation is not needed for the plasma-mode analysis above; it should
be viewed instead as a possible structural imprint of the same hidden
magnetic sector.
 
Several extensions are immediate.  
Magnetized backgrounds should reveal the dual analogues of cyclotron
structure, Faraday rotation, and anisotropic wave propagation.  
Warm, kinetic, or dissipative treatments are needed to describe Landau damping, collisional broadening, and the relaxation spectra expected in effective monopole media such as spin ice.  
The same effective two-fluid equations can also be applied to relativistic
or cosmological settings, including neutral monopole plasmas, hidden-sector
charged media, and the propagation of primordial electromagnetic fields.
Finally, the magnetically neutral dyonic interpretation mentioned above
suggests a separate quantum problem.  
One should classify the admissible composites using the Dirac--Schwinger--Zwanziger charge lattice, including possible Montonen--Olive/Witten-type structures, and then analyze the binding dynamics of the corresponding dyonic states.  
These questions are not needed for the linear-response results derived here, and we leave them to future work.


\end{document}